\documentstyle[11pt,fleqn]{article}
\catcode`\@=11
\@addtoreset{equation}{section}
\def\theequation{\arabic{section}.\arabic{equation}}
\def\appendix{\renewcommand{\thesection}{\Alph{section}}\setcounter{section}{0}
              \renewcommand{\theequation}
            {\mbox{\Alph{section}.\arabic{equation}}}\setcounter{equation}{0}}

\renewcommand{\title}[1]{\begin{center}\Large\bf #1\end{center}\rm\par\bigskip}
\renewcommand{\author}[1]{\begin{center}\Large #1\end{center}}

\newcommand{\email}[1]{e-mail: \sl #1@science.unitn.it\rm}
\newcommand{\femail}[1]{\footnote{\email{#1}}}

\def\babs{\hrule\par\begin{description}\item{Abstract: }\it}
\def\eabs{\par\end{description}\hrule\par\medskip\rm}
\renewcommand{\date}[1]{\par\bigskip\par\sl\hfill #1\par\medskip\par\rm}

\renewcommand{\ss}[1]{\subsection{#1}}

\def\hs{\qquad\qquad}         
\def\nn{\nonumber}            
\def\beq{\begin{eqnarray}}    
\def\eeq{\end{eqnarray}}      
\def\R{{\hbox{{\rm I}\kern-.2em\hbox{\rm R}}}}   
\def\H{{\hbox{{\rm I}\kern-.2em\hbox{\rm H}}}}   
\def\N{{\hbox{{\rm I}\kern-.2em\hbox{\rm N}}}}   
\def\C{{\ \hbox{{\rm I}\kern-.6em\hbox{\bf C}}}} 
\def\Z{{\hbox{{\rm Z}\kern-.4em\hbox{\rm Z}}}}   
\def\ii{\infty}                                  
\renewcommand{\Re}{\,\mbox{Re}\,}       

\def\be{\beta}

\def\de{\delta}
\def\ep{\varepsilon}

\def\si{\sigma}
\def\om{\omega}

\def\th{\theta}

\begin{document}

\title{Radiation from the extremal black holes}

\author{L. Vanzo \femail{vanzo}}

\begin{abstract}
The radiation from extreme Reissner-Nordstr\"{o}m black holes
is computed by explicitly considering the collapse of a spherical charged
shell. No neutral scalar radiation is found but there is emission of
charged particles, provided the charge to mass ratio be different from
one. The absence of thermal effects is in accord with the
predictions of the euclidean theory but since the body emits charged
particles the entropy issue is not the same as for eternal extreme
black holes.
\end{abstract}

\bigskip

\ss{The shell equations}

As was explained in \cite{boul73-8-2363} (see also Ref.s
\cite{isra67-51-744,beke71-4-2185,chas70-67-136}), the
equations of motion of a
collapsing charged spherical shell can be deduced by relating the
discontinuity of the metric across the shell with the surface energy
density in the shell. For an extremal shell, let ${\cal M}$ be the
proper mass and $M$ the total gravitational mass of the shell and
let us introduce the abbreviation
\beq
a=\frac{{\cal M}^2-M^2}{2{\cal M}}\nn
\eeq
The exterior metric is the extremal Reissner-Nordstr\"{o}m metric
\beq
ds^2=-\frac{(r-M)^2}{r^2}dt^2+\frac{r^2}{(r-M)^2}dr^2+r^2d\ell^2
\eeq
and the interior is flat Minkowski space.
The shell position is specified by $r=R(\tau)$, $t=T(\tau)$, where
$\tau$ is the proper time along the shell history.
The equations of motion are then
\beq
\dot{R}=-\sqrt{\left[\frac{M}{{\cal
M}}+\frac{a}{R}\right]^2-1}, \hs
\dot{T}=\frac{R}{R-M}\sqrt{1+\frac{R^2\dot{R}^2}{(R-M)^2}}
\eeq
A simple, but important, consequence of the equations is that
if $M>{\cal M}$, then the shell radius has a turning point at
$R_T=(M+{\cal M})/2<M$ which is a minimum radius. Therefore it
reaches the horizon value
$R=M$, and the shell disappears from sight by forming a future event
horizon. However it subsequentely
bounces and expands by forming another asymptotically flat region with
a past horizon, and singularities are not formed.
The important region for us will be the one containing
the future event horizon.
Although the equations of motion are exactly integrables, we only need
the asymptotic expansion of the functions $R(\tau)$ and $T(\tau)$ near
the horizon. More precisely we need the advanced time $v$ of the shell
as a function of the retarded time $u$, namely the relation $v(u)$ near
the horizon to leading non trivial order. We found
\beq
\dot{R}=a_0+a_1\ep+a_2\ep^2+O(\ep^3), \hs \ep=R-M
\eeq
\beq
\dot{T}=b_{-2}\ep^{-2}+b_{-1}\ep^{-1}+b_0+O(\ep)
\eeq
where the exact form of the coefficients is left apart, as the
really important fact is that both $a_0$ and $b_{-2}$ are non vanishing.
The advanced/retarded coordinate times are respectively $t\pm r^*$,
where
\beq
r^*=r+2M\log(r-M)-\frac{M^2}{r-M}
\eeq
shows the characteristic first order pole of extremal black holes.
{}From the equations it now follows that $v(u)$ remains finite on the
horizon while $u$ diverges to infinity. Therefore
\beq
v(u)=v_1-Cu^{-1}+O(1/\log u), \hs u\rightarrow\ii
\eeq
where $v_1$ is the value of advanced time that marks the birth of the
black hole and $C$ is a positive constant depending on $M$ and
${\cal M}$. This remarkable behaviour is to be contrasted with the non
extreme black hole, where $v(u)$ approaches $v_1$ exponentially fast
in retarded time. This can be traced back to the first order pole
singularity that controls the behaviour of $r^*$ near the horizon in
contrast with the logarithmic singularity of the non extreme case.

\ss{Quantum theory}

The asymptotic behaviour just discussed will have the consequence that
the phase of a positive frequency ingoing mode from past infinity will
be red shifted at a much a lower rate than for non extremal black
holes. Indeed, it is pretty straightforward to deduce the asymptotic
form of such a mode near the event horizon, when the black hole is
about to form. Precisely we have: let $P_{j\om}$ be a positive frequency
solution of the charged Klein-Gordon equation in the collapse geometry
of the infalling shell which is ingoing from past infinity and
therefore ($j$ standing for $(l,m)$ from now on)
\beq
g^{ab}(\nabla_a+ieA_a)(\nabla_b+ieA_b)P_{j\om}-m^2P_{j\om,e}=0
\eeq
\beq
P_{j\om,e}=(4\pi\om)^{-1/2}r^{-1}Y_j(\th,\phi)R_{j\om}(r)e^{-i\om v},
\eeq
with $R_{j\om}(r)\simeq1$ for $r\rightarrow\ii$. Then, asymptotically
for large $u$ and $r\rightarrow M$ we have
\beq
P_{j\om,e}\simeq(4\pi\om)^{-1/2}r^{-1}Y_j(\th,\phi)
\exp(i\om Cu^{-1})e^{-i\si et} \hs \si=\frac{Q}{|Q|}
\label{asym}
\eeq
where $e$ is the charge of the field in the units $G=c=\hbar=1$ that
we are using. The quantum field
\beq
\phi=\sum_j\int_0^{\ii}d\om[A_{j\om}P_{j\om,e}+B^{\dag}_{j\om}
P^*_{j\om,-e}]
\eeq
is assumed to be in the $|in>$ vacuum state relative to the above
decomposition.
In order to discuss the particle content at future
infinity, we should expand the in-mode into positive and negative
frequency waves of the kind $\exp(\pm i\om u)$, namely wavelets which
are purely outgoing to infinity. This is a rather delicate matter.
Considering Eq.(\ref{asym}) as an holomorphic function of $u$ in the
lower half complex $u$-plane, we argue that the Fourier expansion of
$P_{j\om}$ can only contains the factors $\exp(-i\om u)$, which just share
with $P_{j\om}$ the same holomorphic property. We immediately conclude
that {\it there is no neutral scalar radiation at infinity from an
extremal black hole} (no negative frequency components). However, for
charged fields things are differents due to the superradiant
phenomenon. Indeed, in the extended Reissner-Nordstr\"{o}m
geometry the radial outgoing
waves from the past horizon have the asymptotic
behaviour\cite{gibb75-44-245}
\beq
\stackrel{\rightarrow}{R}_{j\om}\simeq\left\{\begin{array}{cc}
e^{i(\om-\si e)r^*}+\stackrel{\rightarrow}{A}_{j,e}(\om)e^{-i(\om-\si
e)r^*}&r^*
\rightarrow-\ii \\ \stackrel{\rightarrow}{B}_{j,e}(\om)e^{i\om r^*}&
r^*\rightarrow\ii\end{array}\right.,\hs |\om|\geq m
\label{outg}
\eeq
which defines the reflection and transmission coefficient of the
potential barrier sorrounding the black hole. One defines similar
reflection-transmission coefficients for ingoing waves from past
infinity. One has then the unitarity relations
\beq
\om\stackrel{\rightarrow}{B}_{j,e}(\om)=(\om-\si e)
\stackrel{\leftarrow}{B}_{j,e}(\om), \hs
(\om-\si e)|\stackrel{\leftarrow}{B}_{j,e}(\om)|^2=\om[
1-|\stackrel{\leftarrow}{A}_{j,e}(\om)|^2]
\label{unit}
\eeq
{}From Eq.(\ref{outg}), a positive frequency wave packet at the horizon
containing the factor $\exp(-i\om u)$, arrives at infinity with the time
dependence $\exp[-i(\om-\si e)u]$. This is holomorphic in the lower half
complex
$u$-plane if $\om-\si e>0$, but is holomorphic in the upper half
complex $u$-plane if $\om-\si e<0$, which is just one super radiance condition.
Therefore the wave packet could have  a
negative frequency component at infinity which just corresponds
to particle emission in the quantum theory.

We now show that this is actually the case. Let $\nu\in\C$ such that
$-1<\Re\nu<1/2$ and $J_{\nu}(z)$ the Bessel function of order $\nu$. Then
\beq
\int_0^{\ii}e^{-i\om^{'}u}(\om^{'})^{\nu/2}J_{\nu}(2\sqrt{\om\om^{'}})
d\om^{'}=\frac{\om^{\nu/2}e^{-i\pi\nu/2}}{iu^{\nu+1}}\exp\left(
\frac{i\om}{u}\right)\th(u)
\eeq
By forming a wave packet one can easily shows that the limit
$\nu\rightarrow-1$ can be safety taken. This means that in the
distributional sense we have
\beq
\int_{-\ii}^{\ii}e^{-i\om^{'}u}(\om^{'})^{-1/2}J_{-1}(2\sqrt{\om\om^{'}})
\th(\om^{'})d\om^{'}=\om^{-1/2}\exp\left(\frac{i\om}{u}\right)\th(u)
\label{dist}
\eeq
This equation, combined with Eq.(\ref{outg}), can be used to propagate
the in-mode $P_{j\om,e}$ to future infinity.
Indeed, each monochromatic component in Eq.(\ref{dist})
will tunnel through the potential barrier sorrounding
the black hole, by carrying the
amplitude $\stackrel{\rightarrow}{B}_{j,e}(\om^{'})$ to infinity (see
Eq.(\ref{outg})). We found
\begin{eqnarray}
P_{j\om,e}&\simeq
&\frac{1}{\sqrt{4\pi}}\int_0^{\ii}e^{-i\om^{'}u}
J_{-1}\left(2\sqrt{\om(\om^{'}-\si
e)}\right)\stackrel{\rightarrow}{B}_{j,e}(\om^{'})
\frac{\th(\om^{'}-\si e)d\om^{'}}{\sqrt{(\om^{'}-\si e)}}\nn \\
&+&\frac{1}{\sqrt{4\pi}}\int_0^{\ii}e^{i\om^{'}u}
J_{-1}\left(2\sqrt{\om(-\si e-\om^{'})}\right)
\stackrel{\rightarrow}{B}^*_{j,-e}(\om^{'})
\frac{\th(-\si e-\om^{'})d\om^{'}}{\sqrt{(-\si e-\om^{'})}}
\label{afie}
\end{eqnarray}
and we can see the presence of a negative frequency component,
provided $\si e<0$. Similarly, a negative frequency in-mode will have
a positive frequency component at infinity if $\si e>0$.
By following the usual procedure \cite{hawk75-43-199}, the Bogoljubov
coefficient
responsible for (charge $e$) particle creation can be found
easily from Eq.(\ref{afie})
\begin{eqnarray}
\be_{j\om^{'}\om,e}&=&-\frac{\sqrt{\om^{'}}}{\sqrt{\si e-\om^{'}}}J_{-1}
\left(2\sqrt{\om(\si e-\om^{'})}\right)
\stackrel{\rightarrow}{B}^*_{j,e}(\om^{'})\th(\si e-\om^{'})\nn \\
&=&\frac{\sqrt{\si e-\om^{'}}}{\sqrt{\om^{'}}}J_{-1}\left(
2\sqrt{\om(\si e-\om^{'})}\right)\stackrel{\leftarrow}{B}^*_{j,e}(\om^{'})
\th(\si e-\om^{'})
\end{eqnarray}
on using Eq.(\ref{unit}). Since $\si$ is the sign of $Q$, the black
hole will only emit
particles having the same sign as $Q$ and the black hole will tend to
discharge. Moreover one has
\beq
\int_0^{\ii}\be^*_{j\om^{'}\om,e}\be_{j\om^{"}\om,e}d\om=-
[1-|\stackrel{\leftarrow}{A}_{j,e}(\om^{'})|^2]\th(pe-\om^{'})
\de(\om^{'}-\om^{"})
\eeq
by using Eq.(\ref{unit}) and the completeness relation for Bessel
functions. From this we obtain the luminosity of the hole as
\beq
L=-\frac{1}{2\pi}\sum_j\int_{m}^{pe}[1-|
\stackrel{\leftarrow}{A}_{j,e}(\om)|^2]\om d\om
\eeq
and the rate of charge loss as
\beq
\dot{Q}=e\sum_j\int_{m}^{pe}[1-|
\stackrel{\leftarrow}{A}_{j,e}(\om)|^2]d\om
\eeq
Remarkably, this is the formula
which holds for non extreme black holes in the limit of zero surface
gravity, regardless the lack of continuity in the asymptotic form of
the solutions. The thermodynamic potentials of black holes are
indeed continuous functions of the
black hole parameters, nevertheless the entropy is discontinuous in
their extreme limit \cite{hawk95-51-4302}, as extreme black holes have
zero entropy but non zero horizon area. Continuity arguments were also
used in \cite{wilc91-6-2353} to argue the break down of thermodynamics
description of extreme black holes. We point out that the luminosity of
charged extreme collapsing shells is the zero temperature limit of the
luminosity formula for non extreme black hole, and in this sense
their temperature vanishes. However, due to emission of charged
particle these objects evolve more likely toward the non extreme
solutions and in this sense their entropy raises to non zero values.

Another interesting observation is that
the luminosity and charge loss both vanish for particles with charge
to mass ratio is unity. So the only possibility for an extremal black hole
to be stable under quantum emission seems to be the non existence of
particles with $e/m\neq 1$. Therefore the black hole will evolve
toward the non extreme sequence, and the issue of extreme black
hole entropy looses its meaning in the present contest. It is only for
eternal extreme black holes that the issue remains
\cite{hawk95-51-4302,mitr94-73-2521}. On the other hand,
the non extreme charged shell almost invariably form a singularity and
there is both thermal emission and geometric entropy. The entropy
problem may thus be connected with the formation of singularities,
rather than with the formation of horizons.


\begin{thebibliography}{1}

\bibitem{boul73-8-2363}
{D.G.~Boulware}.
 Phys.~Rev. {\bf {D8}}, {2363} (1973).

\bibitem{isra67-51-744}
{W.~Israel and V. de la Cruz}.
 Nuovo Cimento {\bf {51}}, {774} (1967).

\bibitem{beke71-4-2185}
{J.D.~Bekenstein}.
 Phys.~Rev. {\bf {D4}}, {2185} (1971).

\bibitem{chas70-67-136}
{J.E. Chase}.
 Nuovo Cimento {\bf {67B}}, {136} (1970).

\bibitem{gibb75-44-245}
{G.W.~Gibbons}.
 Commun.~Math.~Phys. {\bf {44}}, {265} (1975).

\bibitem{hawk75-43-199}
S.W.~Hawking.
 Commun.~Math.~Phys. {\bf 43}, 199 (1975).

\bibitem{hawk95-51-4302}
S.W.~Hawking, G.~Horowitz and S.~Ross.
 Phys.~Rev. {\bf D 51}, 4302 (1995).

\bibitem{wilc91-6-2353}
{J.~Preskill, P.~Schwarz, A.~Shapere, S.~Trivedi and F.~Wilczek}.
 {Mod. Phys. Lett. A} {\bf {6}}, {2353} (1991).

\bibitem{mitr94-73-2521}
{A.~Ghosh and P.~Mitra}.
 Phys.~Rev.~Lett. {\bf {73}}, {2521} (1994).

\end{thebibliography}
\end{document}